\documentclass[12pt,a4paper]{article}
\usepackage{amsmath}
\usepackage{amssymb}
\usepackage{bm}
\usepackage{color}

\usepackage{hyperref}
\usepackage{graphicx}

\catcode `@=11 \@addtoreset{equation}{section} \catcode `@=12

\begin{document}
\setlength{\oddsidemargin}{0cm}
\setlength{\baselineskip}{7mm}

\begin{titlepage}

	\begin{center}
		{\LARGE
			Extracting classical Lyapunov exponent \\
			 from one-dimensional quantum mechanics
        }
	\end{center}
	\vspace{0.2cm}
	\baselineskip 18pt 
	\renewcommand{\thefootnote}{\fnsymbol{footnote}}

	\begin{center}
		Takeshi {\sc Morita}\footnote{%
			E-mail address: morita.takeshi@shizuoka.ac.jp
		}
		
		\renewcommand{\thefootnote}{\arabic{footnote}}
		\setcounter{footnote}{0}
		
		\vspace{0.4cm}
		
		{\small\it

			Department of Physics,
			Shizuoka University, \\
			836 Ohya, Suruga-ku, Shizuoka 422-8529, Japan\\
			
			Graduate School of Science and Technology, Shizuoka University,\\
			836 Ohya, Suruga-ku, Shizuoka 422-8529, Japan
			
		}

	\end{center}
	
	
	\vspace{1.5cm}
	
	\begin{abstract}
        The commutator $[x(t),p]$ in an inverted harmonic oscillator (IHO) in one-dimensional quantum mechanics exhibits remarkable properties. 
		It reduces to a c-number and does not show any quantum fluctuations for arbitrary states.
		Related to this nature, the quantum Lyapunov exponent computed through the out-of-time-order correlator (OTOC) $\langle [x(t),p]^2 \rangle $ precisely agrees with the classical one.
		Hence, the OTOC may be regarded as an ideal indicator of the butterfly effect in the IHO. 
		Since IHOs are ubiquitous in physics, these properties of the commutator $[x(t),p]$ and the OTOCs might be seen in various situations, too.
		In order to clarify this point, as a first step, we investigate OTOCs in one-dimensional quantum mechanics with polynomial potentials, which exhibit butterfly effects around the peak of the potential in classical mechanics.
		We find two situations in which the OTOCs show exponential growth reproducing the classical Lyapunov exponent of the peak.
		The first one, which is obvious, is using a suitably localized wave packet near the peak, and the second one is taking a limit akin to the large-$N$ limit in the noncritical string theories.
		
	\end{abstract}
	
	
\end{titlepage}

\section{Introduction}

Inverted harmonic oscillators (IHOs) are ubiquitous in nature.
If we drew a random potential on paper, we would see as many peaks as valleys.
The valleys will be approximated by harmonic oscillators (HOs), and the peaks will be approximated by IHOs.
Needless to say, HOs play indispensable roles in physics, particularly in stable systems.
Correspondingly, IHOs play crucial roles in unstable systems.

In particular, IHOs appear in several important topics in modern physics:
dynamical systems and chaos\footnote{
Particle motion in an exact IHO \eqref{IHO} is integrable and not chaotic.
However, in typical chaotic systems, the appearance of their chaotic behaviors is explained through motions near hyperbolic fixed points with broken homoclinic orbits \cite{Wiggins:215055}, and the hyperbolic fixed points are approximately described by IHOs.
Thus, IHOs capture several essential properties of chaos. This is one motivation of this work.
 } \cite{Wiggins:215055},
Schwinger mechanisms \cite{BALAZS1990123, PARENTANI1992474, Brout:1995rd}, noncritical string theories \cite{Klebanov:1991qa,Ginsparg:1993is,Polchinski:1994mb, Nakayama:2004vk}, toy models of black holes \cite{Karczmarek:2004bw, Banks:2015mya, Hashimoto:2016dfz, Betzios:2016yaq, Maitra:2019eix, Dalui:2019esx, Dalui:2020qpt, Majhi:2021bwo}, acoustic Hawking radiation in quantum fluid mechanics \cite{Giovanazzi:2004zv, Giovanazzi:2006nd, Parola:2017qpn, Morita:2018sen, Morita:2019bfr, Morita:2021mfi}, and condensed matter systems \cite{PhysRevA.82.053617, Hegde:2018xub, Kelly:2020vty, Subramanyan:2020fmx}.

These examples show some instabilities, and they may be quantified by Lyapunov exponents at the classical level.
Recently, as a counterpart to this quantity in quantum mechanics, the out-of-time-order correlator (OTOC) \cite{1969JETP...28.1200L} defined by
\begin{align}
	C(t):=-\langle [W(t),  V(0)]^2 \rangle 
	\label{OTOC-def}
\end{align}
has attracted attention \cite{Kitaev:2015, Kitaev:2015-2, Maldacena:2015waa}.
Here, we use the Heisenberg picture, $W$ and $V$ are some operators in the system, and $W(t)=e^{iHt/\hbar}W(0) e^{-iHt/\hbar} $ by using the Hamiltonian $H$.
If we take $W=x$ and $V=p$ in a quantum mechanical system, 
Eq.~\eqref{OTOC-def} becomes
\begin{align}
	- \frac{1}{\hbar^2} \langle [x(t),  p(0)]^2 \rangle \to  \{ x(t),p(0) \}^2 =\left(\frac{\partial x(t)}{\partial x(0)} \right)^2  ,
	\label{OTOC-XP}
\end{align}
where we have used the classical-quantum correspondence, $[,]/i\hbar \to \{,\} $.
Thus, the OTOC evaluates the dependence of the initial condition of the time evolutions.
Particularly, if the system shows a butterfly effect at the classical level, the OTOC may develop as $C(t) \sim e^{2\lambda t} $, 
where $\lambda$ is the Lyapunov exponent.
Hence, OTOCs may quantify butterfly effects in quantum mechanics.\footnote{In this article, the terminology ``butterfly effect" is used for the sensitive dependence of the initial condition, i.e., a finite positive Lyapunov exponent. 
Note that the sensitive dependence is usually related to some instabilities of the system, and it occurs even in nonchaotic systems.
}

However, the relation between OTOCs and Lyapunov exponents is subtle.
First, we assume the classical-quantum correspondence, and it does not work, in general.
Second, even if the classical-quantum correspondence is satisfied at the early stage of the time evolution, it may break down after the Ehrenfest time and the exponential development may not be observed after that.
Thus, detecting exponential developments in quantum systems is harder than in the classical ones \cite{Maldacena:2015waa}.
Hence, it is valuable to understand when we observe $C(t) \sim  e^{2\lambda t}$ in order to reveal properties of the OTOCs.
Since IHOs also show butterfly effects, it is natural to investigate the OTOCs in IHOs in detail and study the application to the aforementioned systems.

In this article, for simplicity, we consider the IHO in one-dimensional quantum mechanics \cite{Hashimoto:2016wme, Xu:2019lhc, Bhattacharyya:2020art, Hashimoto:2020xfr},
\begin{align}
	H=\frac{1}{2}p^2-\frac{1}{2}\lambda^2 x^2 .
	\label{IHO}
\end{align}
Here, $\lambda$ is the Lyapunov exponent, as we will see soon.
The quantum Lyapunov exponent in the IHO has been computed by evaluating an OTOC in Ref.~\cite{Hashimoto:2016wme}, and it exactly agrees with the classical one.
Particularly, the results of Ref.~\cite{Hashimoto:2016wme} imply the following relation,
\begin{align}
	\left\langle \left( \frac{1}{i\hbar}  [x(t),p(0)] \right)^n  \right\rangle & = \cosh^n \lambda t =  \left(  \{ x(t),p(0) \} \right)^n .
	\label{OTOC-proposal}
\end{align}
Here, the left-hand side evaluates the OTOC in the IHO \eqref{IHO} for any normalizable quantum states, and the right-hand side is the Poisson bracket for any initial conditions in classical mechanics.\footnote{In Eq.~\eqref{OTOC-proposal}, the $n=1$ case in the left-hand side may not be suitable to be called an OTOC. However, Eq.~\eqref{OTOC-proposal} shows the exponential development at large $t$ and diagnoses the butterfly effect, and we loosely call it an OTOC in this article. }
Since $\cosh(\lambda t) \sim e^{\lambda t}$ at a large $t$, this relation shows that  the Lyapunov exponent of this system is $\lambda$ in both classical and quantum mechanics, as Ref.~\cite{Hashimoto:2016wme} found.
Note that this relation works for any time even after the Ehrenfest time, which is typically given by $t \sim \frac{1}{\lambda} \log \frac{1}{\hbar}$ \cite{Maldacena:2015waa}.\footnote{The Ehrenfest time $t \sim \frac{1}{\lambda} \log \frac{1}{\hbar}$ is estimated as the timescale that a wave packet spreads over the curvature scale of the IHO.
However, the domain of the IHO \eqref{IHO} is infinite ($-\infty \le x \le \infty$), and it may be reasonable that the naive Ehrenfest time does not work in our case.
}
Furthermore, this relation suggests that the commutator $[x(t),p(0)]$ does not show any quantum fluctuations and the deviation is precisely zero.
Thus, the commutator $[x(t),p(0)]$ exhibits quite peculiar properties in the IHO, which cannot be seen in other observables, like $x(t)$ and $p(t)$.
These results suggest that the commutator $[x(t),p(0)]$ may be regarded as an ideal indicator of the butterfly effect in the IHO.

Then, it is natural to ask whether these remarkable properties of the OTOCs hold in more general situations.
In order to understand this question, we study one-dimensional quantum mechanics with polynomial potentials \cite{Hashimoto:2020xfr, Romatschke:2020qfr}.
In classical mechanics, the particle motions confined in the potentials are periodic and not chaotic.
However, if the potential has a hill, the hill will be approximated by an IHO, and the system shows a butterfly effect near there. 
We discuss when the OTOCs reproduce this classical Lyapunov exponent of the hill in quantum mechanics.
As is expected through the classical-quantum correspondence, we see that suitably localized wave packets correctly reproduce the Lyapunov exponent. 
In addition, if we take a limit similar to the large-$N$ limit in the noncritical string theories \cite{Klebanov:1991qa,Ginsparg:1993is,Polchinski:1994mb, Nakayama:2004vk}, the correct Lyapunov exponent will be obtained through more general states such as energy eigenstates.

The organization of this article is as follows.
In Sec.~\ref{sec-IHO}, we study the nature of the OTOCs in the IHO \eqref{IHO} in detail.
In Sec.~\ref{sec-general}, we investigate the OTOCs in more general potential cases.
In Sec.~\ref{sec-classical}, we argue that one can understand our results in the $\hbar \to 0$ limit in
 terms of classical mechanics.
Section \ref{sec-discussion} contains conclusions and discussions.

\section{No Quantum Fluctuation of the OTOC in the IHO}
\label{sec-IHO}
We prove the relation \eqref{OTOC-proposal}.
We start from classical mechanics.
The classical solution of the Hamiltonian \eqref{IHO} is given by
\begin{align}
	x(t)=x(0) \cosh \lambda t + \frac{1}{\lambda} p(0) \sinh \lambda t,
	\label{classical-motion}
\end{align}
where $x(0)$ and $p(0)$ are the initial conditions of the position $x(t)$ and momentum $p(t)$.
Then, we can compute the Poisson bracket as
\begin{align}
	\{ x(t),p(0) \} = \frac{\partial x(t)}{\partial x(0)} = \cosh \lambda t,
\end{align}
and the second equality in the relation \eqref{OTOC-proposal} is satisfied.

Next, we consider quantum mechanics.
As Refs.~\cite{Hashimoto:2016wme, Bhattacharyya:2020art} pointed out, we obtain 
\begin{align}
	x(t) = e^{iHt} x(0) e^{-iHt} = x(0) \cosh \lambda t + \frac{1}{\lambda} p(0) \sinh \lambda t 
\end{align}
through the Hadamard lemma, and it leads to
\begin{align}
	[x(t),p(0)] = i \hbar \cosh \lambda t .
\end{align}
Since this quantity is a c-number, the relation \eqref{OTOC-proposal} is satisfied for any normalizable states.
Obviously, similar relations hold for other commutators, $[p(t),p(0)]$, $[x(t),x(0)]$, and $[p(t),x(0)]$.

\begin{figure}
	\begin{tabular}{ccc}
		\begin{minipage}{0.33\hsize}
			\begin{center}
				\includegraphics[scale=0.4]{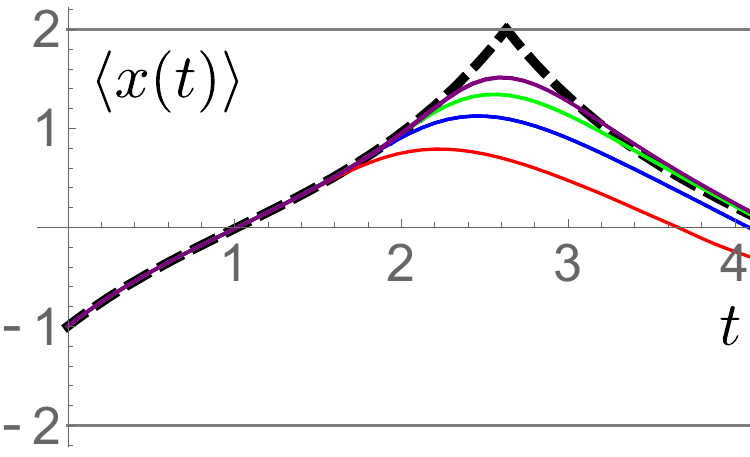}
			\end{center}
		\end{minipage}
		\begin{minipage}{0.33\hsize}
			\begin{center}
				\includegraphics[scale=0.4]{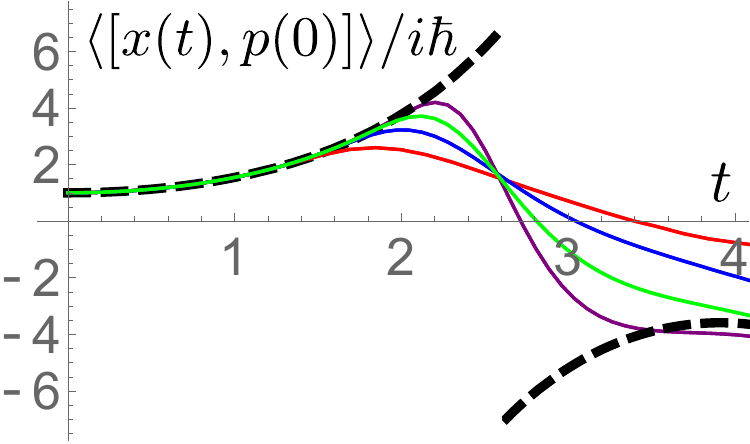}
			\end{center}
		\end{minipage}
		\begin{minipage}{0.33\hsize}
			\begin{center}
				\includegraphics[scale=0.4]{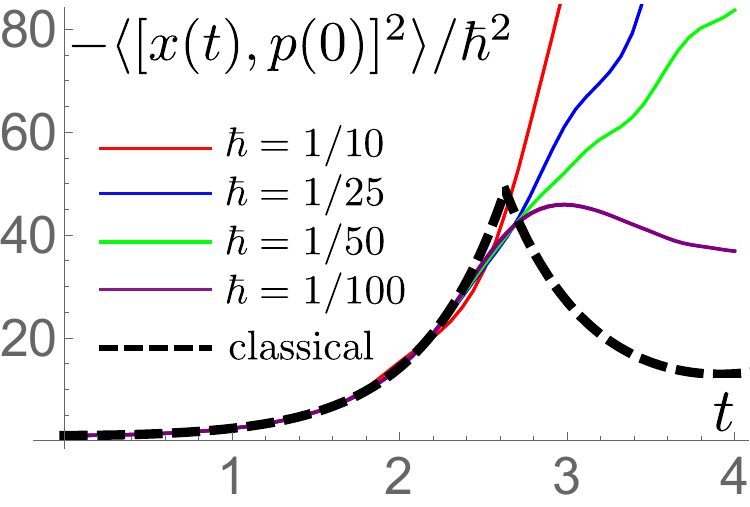}
			\end{center}
		\end{minipage}
	\end{tabular}
	\caption{
		Time evolutions of $x(t)$ and the OTOCs in the IHO \eqref{IHO}.
		We take $\lambda=1$ and put the infinite potential walls at $x= \pm 2$.
		We prepare the Gaussian wave packets with $(\Delta x)^2=(\Delta p)^2$ centered at $(x,p)=(-1,1.3)$ at $t=0$ and evaluate their time evolutions 
		in the quantum mechanics with $\hbar=1/10, 1/25, 1/50$, and $1/100$.
		We also compute the corresponding quantities for a single classical particle; they are depicted by the black dashed lines.
		Note that $\langle x(t) \rangle$ shows that the wave packets hit the potential wall at $x=2$ around $t\sim 2.5$.
		All the OTOCs agree very well until the hits, and they are independent of $\hbar$.
		Thus, the relation \eqref{OTOC-proposal} works as long as we ignore the effect of the potential walls.
		}
	\label{Fig-wave}
\end{figure}

However, the result \eqref{OTOC-proposal} is subtle since the IHO potential \eqref{IHO} is unbounded from below and the energy eigenfunctions of the IHO \eqref{IHO} are not normalizable.
Hence, it would be valuable to test the relation \eqref{OTOC-proposal} explicitly.
We add the infinite potential walls at $x=\pm \Lambda$ to the IHO \eqref{IHO} to make the system bounded and compute the OTOC 
$\left\langle  [x(t),p(0)]  \right\rangle $ and $\left\langle  [x(t),p(0)]^2  \right\rangle $ for several Gaussian wave packets numerically.\footnote{We use {\it Mathematica} package NDEigensystem in the numerical calculations.
However, this package sometimes fails to obtain eigenfunctions that have suitable parity symmetry $x \to -x$.
Hence, we take the domain $0 \le x \le \Lambda $ rather than $-\Lambda  \le x \le \Lambda $ and solve the even and odd solutions separately by imposing the corresponding boundary conditions at $x=0$.  
}
We find very good agreement between classical and quantum mechanics as shown in Fig.~\ref{Fig-wave}, and the quantum fluctuations of the operator $  [x(t),p(0)]  $
are indeed almost zero.
Thus, our numerical computations verify the prediction \eqref{OTOC-proposal}.

Note that if we replace $\lambda$ with $i \omega$, $(\omega \in {\mathbf R})$ in the Hamiltonian \eqref{IHO}, we obtain a similar relation for the harmonic oscillator (HO),
\begin{align}
	\left\langle \left( \frac{1}{i\hbar}  [x(t),p(0)] \right)^n  \right\rangle & = \cos^n \omega t =  \left(  \{ x(t),p(0) \} \right)^n 
	\label{OTOC-HO}.
\end{align}
Actually, we can derive this relation directly from the relation $\langle m| [x(t),p(0)] |n\rangle = i \hbar \delta_{mn} \cos \omega t$, which we can easily obtain by solving the HO through the standard method \cite{Hashimoto:2017oit}.

\section{OTOC in General Potentials}
\label{sec-general}

So far, we have seen that the quantum fluctuations of the OTOCs in the IHO and HO are exactly zero.
This is because $x(t)$ in Eq.~\eqref{classical-motion} is linear in $x(0)$ and $p(0)$, and it will not be true in the general potential $V(x)$.
On the other hand, if the potential $V(x)$ has a hill (valley), the region near the hill (valley) will be approximated by the IHO (HO), and the quantum fluctuations of the OTOCs will be suppressed.
Particularly, a classical particle near the hill will show a butterfly effect, and the Lyapunov exponent is computed from the curvature of the potential as
\begin{align}
	\lambda_{\rm saddle}:= \left. \sqrt{-V''(x)} \right|_{x=x_{\rm saddle}}.
\end{align}
Here, we have taken $x=x_{\rm saddle}$ as the position of the top of the hill since it is a saddle point in phase space, and we have defined $\lambda_{\rm saddle}$ as the Lyapunov exponent associated with this point.
Hence, if we can prepare sufficiently localized wave packets corresponding to classical particles near the hill, the OTOCs would show the exponential developments with the Lyapunov exponent $\lambda_{\rm saddle}$, and the quantum corrections would be small.
(In order to prepare such localized wave packets in the general potential $V(x)$, the potential hill should be sufficiently isolated.) 
Indeed, we have seen in Fig.~\ref{Fig-wave} that the deformation of the IHO potential by the infinite walls does not affect the relation \eqref{OTOC-proposal}.

\begin{figure}
	\begin{tabular}{c||c}
		$V(x)=-a x^2 +b x^4$ &
		$V(x)=-a x^2 +b x^8$ \\
		\begin{minipage}{0.5\hsize}
			\begin{center}
				\includegraphics[scale=0.4]{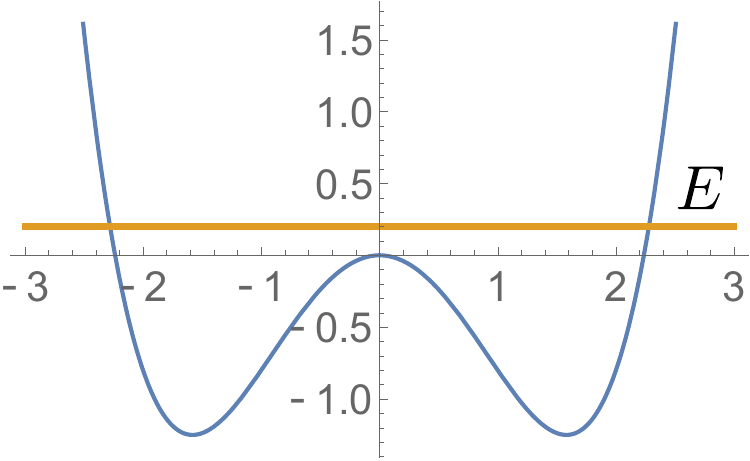}
			\end{center}
		\end{minipage}
		&
		\begin{minipage}{0.5\hsize}
			\begin{center}
				\includegraphics[scale=0.4]{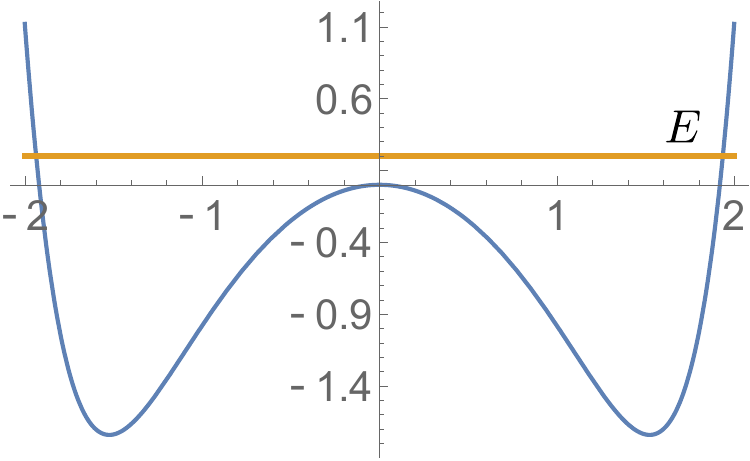}
			\end{center}
		\end{minipage} \\
		\hline
		\begin{minipage}{0.5\hsize}
			\begin{center}
				\vspace{5pt}\hspace{-70pt}$ \langle [x(t),p(0)] \rangle /i \hbar $\\
				\includegraphics[scale=0.4]{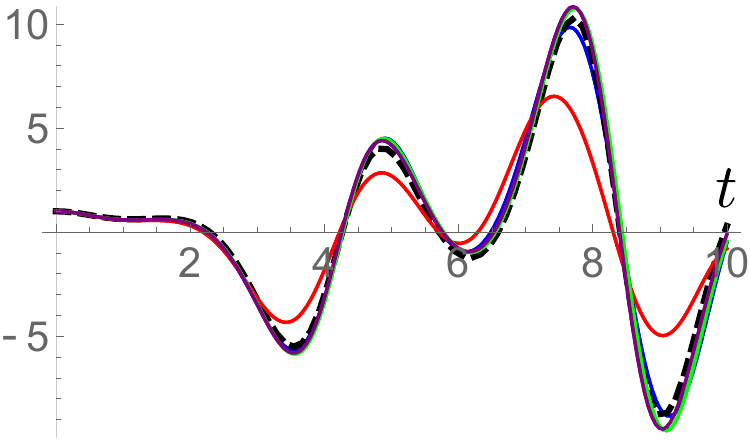}
			\end{center}
		\end{minipage}
		&
		\begin{minipage}{0.5\hsize}
			\begin{center}
				\vspace{5pt}\hspace{-60pt}$ \langle [x(t),p(0)] \rangle /i \hbar $\\
				\includegraphics[scale=0.4]{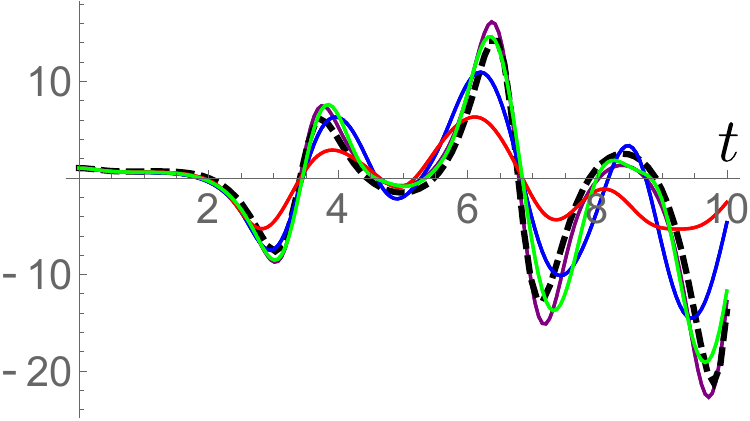}
			\end{center}
		\end{minipage} \\
		\hline		
		\begin{minipage}{0.5\hsize}
			\begin{center}
				\vspace{5pt}\hspace{-60pt}$-\langle [x(t),p(0)]^2 \rangle / \hbar^2  $\\
				\includegraphics[scale=0.4]{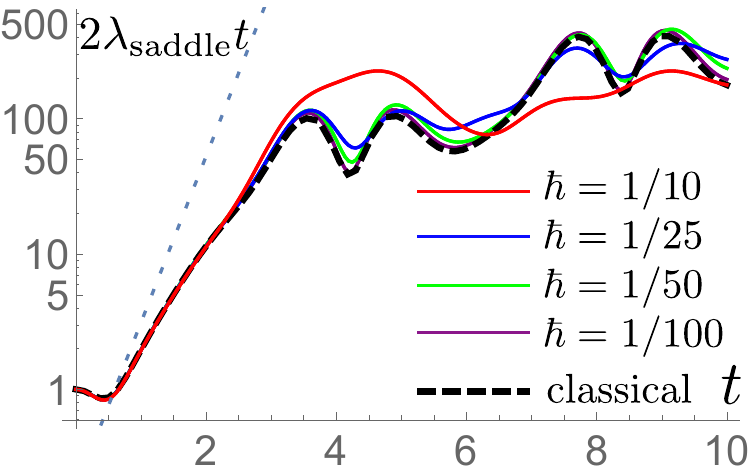}
			\end{center}
		\end{minipage}
		&
		\begin{minipage}{0.5\hsize}
			\begin{center}
				\vspace{5pt}\hspace{-60pt}$-\langle [x(t),p(0)]^2 \rangle / \hbar^2 $\\
				\includegraphics[scale=0.4]{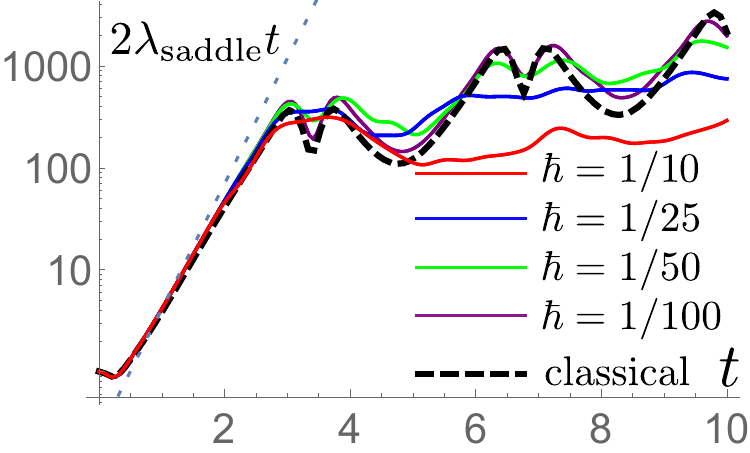}
			\end{center}
		\end{minipage} 
	\end{tabular}
	\caption{
	OTOCs for energy eigenstates in $V(x)=-ax^2+bx^{4}$ ($a=1$, $b=0.2$) (left) and $V(x)=-ax^2+bx^{8}$ ($a=1$, $b=0.02$) (right).
	The energies are taken to be $E=0.2$ in both cases.
	(In quantum mechanics, the energy eigenstates whose energy is closest to $E=0.2$ are taken.)
	 We evaluate $\langle [x(t),p(0)] \rangle $ and $\langle [x(t),p(0)]^2 \rangle $, and their classical counterparts, which are depicted by the black dashed lines.
	 (The derivation of the classical results is explained in Sec.~\ref{sec-classical}.)
	 As $\hbar \to 0 $, the OTOCs converge to the classical results.
	 We do not observe clear exponential developments in $\langle [x(t),p(0)] \rangle $.
	 On the other hand, we observe exponential growth in $\langle [x(t),p(0)]^2 \rangle $; however, the exponent in the $V(x)=-ax^2+bx^{4}$ case is significantly smaller than $2\lambda_{\rm saddle}$, while it is close to $2\lambda_{\rm saddle}$ in the  $V(x)=-ax^2+bx^{8}$ case.
	}
	\label{Fig-ene}
\end{figure}

Then, one question is whether one can obtain $\lambda_{\rm saddle}$ through the OTOCs without using the localized wave packets.
Particularly, energy eigenstates are a useful basis of the Hilbert space, and it is natural to try to evaluate the OTOCs for these states.
However, energy eigenstates generally do not represent a localized particle in the position space, and obtaining $\lambda_{\rm saddle}$ from them seems nontrivial.
In order to test it, we numerically evaluate the OTOCs $\langle [x(t),p(0)] \rangle $ and $\langle [x(t),p(0)]^2 \rangle $  for the energy eigenstates in the potential\footnote{
We have attempted to evaluate the OTOCs of the energy eigenstates in the IHO with the infinite potential walls, which we have used in the numerical study of the wave packets in Fig.~\ref{Fig-wave}.
However, we found that the convergence of numerical computations was not good, and we could not obtain reliable results.
We presume that the infinite potential walls are problematic.
Actually, if we evaluate $ \{x(t),p(0)\}^2 $ of a classical particle in an infinite potential well (without the IHO potential) and take an average over the initial position so that it corresponds to the semiclassical energy eigenstate, we can easily see that $ \{x(t),p(0)\}^2 $ diverges.
This divergence will be resolved in quantum mechanics, but it may cause larger numerical errors.
} $V(x)=-ax^2+bx^{4}$ and $V(x)=-ax^2+bx^{8}$, ($a,b>0$).
The results are summarized in Fig.~\ref{Fig-ene}.
We do not observe clear exponential developments for $\langle [x(t),p(0)] \rangle $.
On the other hand, exponential growth is observed for $\langle [x(t),p(0)]^2 \rangle $, but the exponents depend on the shape of the potentials.
Although the exponent in the $V(x)=-ax^2+bx^8$ case is close to $2\lambda_{\rm saddle}$, it is significantly smaller in the $V(x)=-ax^2+bx^4$ case.\footnote{In the $V(x)=-ax^2+bx^4$ case, the exponential development of $\langle [x(t),p(0)]^2 \rangle $ is roughly $\langle [x(t),p(0)]^2 \rangle \sim  \exp   \left(   \lambda_{\rm saddle}t \right) \neq \exp   \left(  2 \lambda_{\rm saddle}t \right) $.
Similar behaviors have been observed in other models too \cite{Xu:2019lhc}, and Ref.~\cite{Xu:2019lhc} argued that the OTOCs may be suppressed by $ \exp   \left(  - \lambda_{\rm saddle}t \right)$ in thermal ensembles.
}
In addition, the results must depend on the energy level. 
For example, if the energy is close to the ground state, the particles are localized near the bottom of the potential, and we would observe the cos-type behaviors as in \eqref{OTOC-HO} rather than exponential developments.

\begin{figure}
	\begin{tabular}{c||c}
		$V(x)=-a x^2 +b x^4$ &
		$V(x)=-a x^2 +b x^8$ \\
		\begin{minipage}{0.5\hsize}
			\begin{center}
				\includegraphics[scale=0.4]{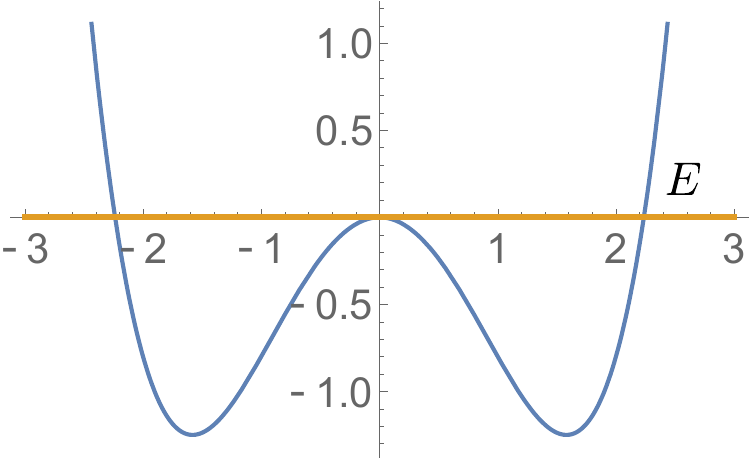}\\
			\end{center}
		\end{minipage}
		&
		\begin{minipage}{0.5\hsize}
			\begin{center}
				\includegraphics[scale=0.4]{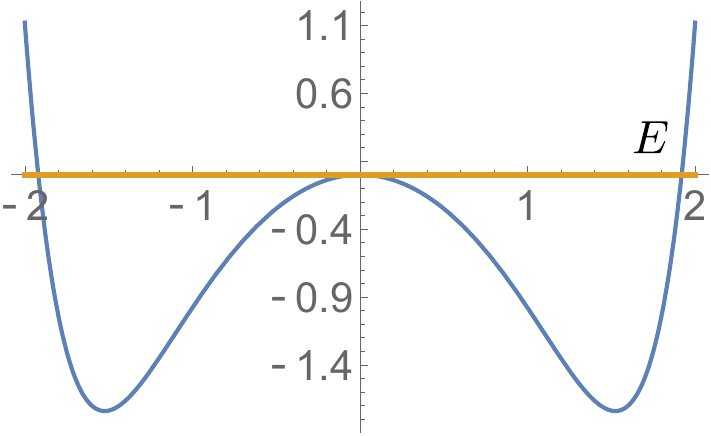}\\
			\end{center}
		\end{minipage} \\
		\hline
		\begin{minipage}{0.5\hsize}
			\begin{center}
				\vspace{5pt}\hspace{-60pt}$\langle [x(t),p(0)] \rangle /i \hbar $\\
				\includegraphics[scale=0.4]{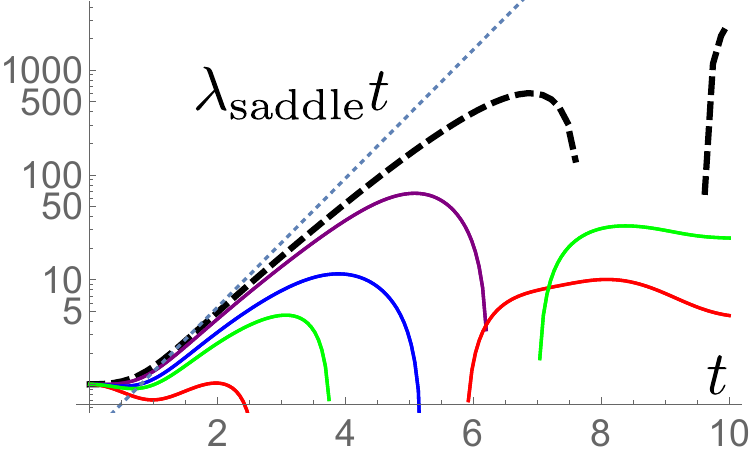}
			\end{center}
		\end{minipage}
		&
		\begin{minipage}{0.5\hsize}
			\begin{center}
				\vspace{5pt}\hspace{-60pt}$\langle [x(t),p(0)] \rangle /i \hbar$\\
				\includegraphics[scale=0.4]{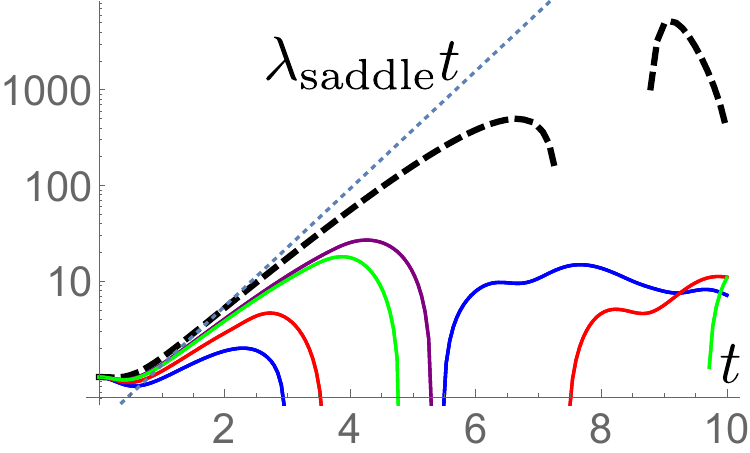}
			\end{center}
		\end{minipage} \\
		\hline		
		\begin{minipage}{0.5\hsize}
			\begin{center}
				\vspace{5pt}\hspace{-60pt}$ -\langle [x(t),p(0)]^2 \rangle / \hbar^2  $\\
				\includegraphics[scale=0.4]{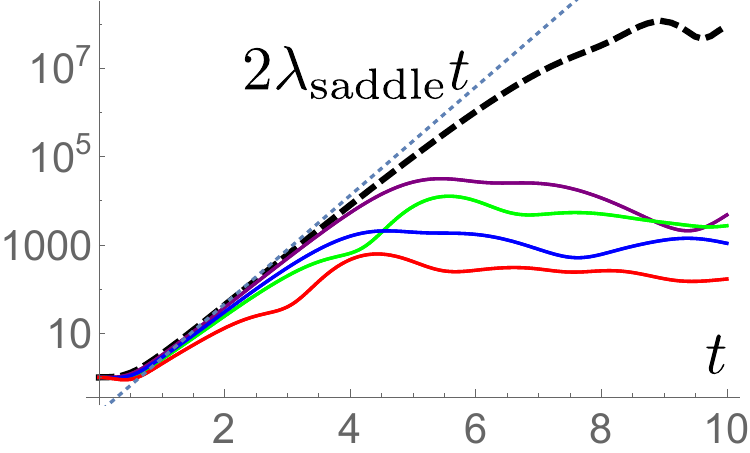}
			\end{center}
		\end{minipage}
		&
		\begin{minipage}{0.5\hsize}
			\begin{center}
				\vspace{5pt}\hspace{-60pt}$-\langle [x(t),p(0)]^2 \rangle / \hbar^2  $\\
				\includegraphics[scale=0.4]{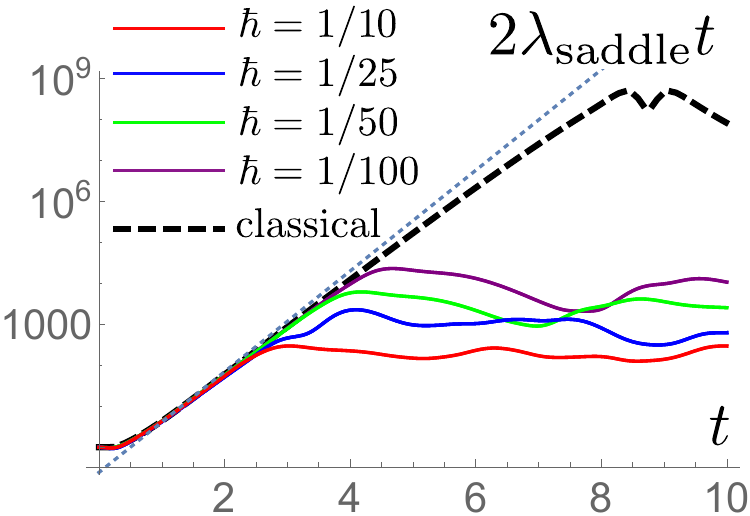}
			\end{center}
		\end{minipage} 
	\end{tabular}
	\caption{
		OTOCs for energy eigenstates close to the critical energy $E_{\rm cr}$ in Eq.~\eqref{E-cr}.
		We employ the same data in Fig.~\ref{Fig-ene} except the energies.
		In quantum mechanics, since we cannot take $E=E_{\rm cr}$ exactly, we choose the closest one with $E-E_{\rm cr}>0$, and we take $E=0.0001$ in classical mechanics, correspondingly.
		We observe that all OTOCs show the exponential growth with the Lyapunov exponent $\lambda_{\rm saddle}$ as $\hbar \to 0$, and the relation \eqref{OTOC-proposal} is approximately satisfied.
	}
	\label{Fig-ene-cr}
\end{figure}

Actually, if we tune the energy level and $\hbar$ such that 
\begin{align} 
	E \to E_{\rm cr}:= V(x_{\rm saddle}), \qquad \hbar \to 0,
	\label{E-cr}
\end{align}
the OTOCs for the energy eigenstate show the exponential developments (see Fig.~\ref{Fig-ene-cr}). 
We observe that both the OTOCs $\langle [x(t),p(0)] \rangle $ and $\langle [x(t),p(0)]^2 \rangle $ show the exponential growth with the Lyapunov exponent $\lambda_{\rm saddle}$ as we take $\hbar \to 0$.
This limit is related to the double-scaling limit in the noncritical string theories \cite{Klebanov:1991qa,Ginsparg:1993is,Polchinski:1994mb, Nakayama:2004vk}, and we call this energy $E_{\rm cr}$ the critical energy.

This result can be explained as follows. 
In the Wentzel–Kramers–Brillouin (WKB) approximation, the probability density of the energy eigenfunction $\rho$ is proportional to the inverse of the classical momentum, $\rho \propto 1/|p|$.
At $E = E_{\rm cr} = V(x_{\rm saddle})$, the momentum near $x=x_{\rm saddle}$ satisfies 
\begin{align}
	0 \sim \frac{1}{2}p^2-\frac{1}{2}\lambda_{\rm saddle}^2 (x-x_{\rm saddle})^2,
	\label{p-critical}
\end{align}
and the density $\rho$ shows a divergence
\begin{align} 
	\rho \propto \frac{1}{|p|} \propto  \frac{1}{|x-x_{\rm saddle}|}.
	\label{rho-critical}
\end{align}
Therefore, when we evaluate observables, the contribution of the saddle points would dominate.\footnote{The momentum becomes zero at the turning point $x=x_*$ also, at which
$E=V(x_*)$ is satisfied.
However, the momentum normally behaves as $|p| \sim (x-x_*)^{1/2} $, and the divergence is much milder than that of the saddle point at the critical energy \eqref{rho-critical}.
Hence, the turning points do not provide dominate contributions.
} 
However, quantum corrections make this divergence milder.
By combining these two effects, the OTOCs exhibit behaviors similar to the IHO near the saddle point as $E \to  E_{\rm cr} $ and $\hbar \to 0$.
Note that, if we prepare some states that are constructed from energy eigenstates whose energies are close to  $E_{\rm cl} $, their OTOCs may also show the exponential growth.\footnote{In noncritical string theories, the ground states of $N$ free fermions in double-well-type potentials with the Fermi energy $E_{\rm F}$ and $\hbar =1/N $ are studied.
This system under the double scaling limit [$\hbar \to  0$ ($N \to \infty $) while taking $E_{\rm cr} - E_{\rm F}$ small but finite] would describe the noncritical string theory \cite{Klebanov:1991qa,Ginsparg:1993is,Polchinski:1994mb, Nakayama:2004vk}.
From our arguments, the fluctuations of the Fermi surface would show the exponential developments of the OTOCs, and it may be valuable to understand the implication in the context of the noncritical string theories.
}

\section{Understanding OTOCs from Classical Mechanics}
\label{sec-classical}

As we can see in Figs.~\ref{Fig-ene} and \ref{Fig-ene-cr}, the OTOCs converge to the classical results as $\hbar \to 0$.
Thus, the behaviors of the OTOCs may be explained in terms of classical mechanics. 
In this section, we discuss the following two questions through classical mechanics: (1) Why are the exponents in Figs.~\ref{Fig-ene} and \ref{Fig-ene-cr} always smaller than $\lambda_{\rm saddle}$ or $2\lambda_{\rm saddle}$. (2) In Fig.~\ref{Fig-ene}, why does $\langle [x(t),p(0)]^2 \rangle $ show the exponential development while $\langle [x(t),p(0)] \rangle $ does not. 

\begin{figure}
	\begin{center}
		\includegraphics[scale=0.4]{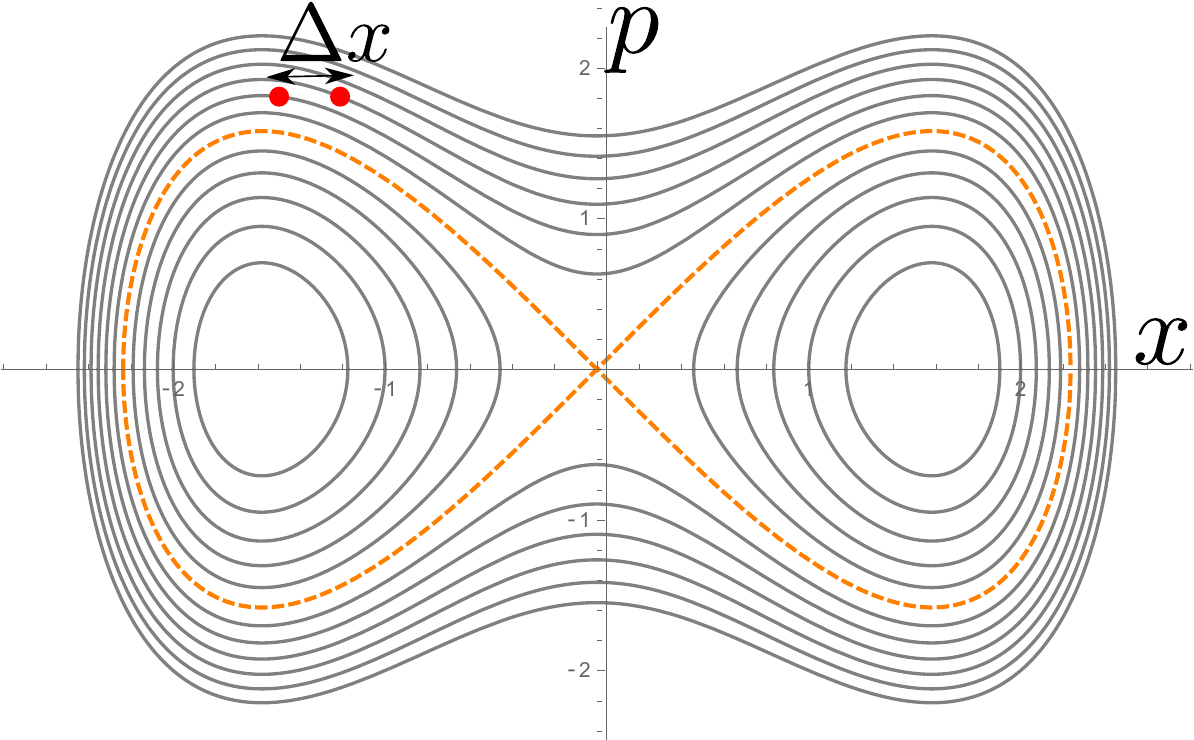}
		\caption{
			Constant energy curves (energy contours) in phase space for the potential $V(x)=-ax^2+bx^{4}$ ($a>0$, $b>0$).
			The orange dashed line denotes the critical energy $E_{\rm cr}$.
			We can compute the Poisson bracket \eqref{otoc-classical} by evaluating the motions of the two particles separated by $\Delta x$ at $t=0$. 
			Each particle orbits in a clockwise direction along the constant energy curve, and the Poisson bracket \eqref{otoc-classical} can be obtained through the deviation of the positions of these particles. 
		}
		\label{Fig-phase}
	\end{center}
\end{figure}

First, we consider  the derivation of the Poisson bracket $\{ x(t),p(0) \}^n$, which is the counterpart of the OTOC $\left\langle  [x(t),p(0)]^n  \right\rangle $.
In classical mechanics, the energy eigenstate in quantum mechanics can be approximated by using the particles uniformly distributed on the constant energy curve in phase space (see Fig.~\ref{Fig-phase}).  
Then, physical quantities for the energy eigenstate can be computed by taking the averages of the quantities for each particle.
Hence, to obtain the OTOC, we need to compute $\{ x(t),p(0) \}^n$ for single particles and take their average.

We can compute  the Poisson bracket $\{ x(t),p(0) \}$ for a single particle as follows.
Suppose that a particle starts from $(x,p)=(x(0),p(0))$ at $t=0$, and we define the position of this particle at time $t$ as $x(t,x(0),p(0))$.
Then, we can compute  
\begin{align}
	\{ x(t),p(0) \} = \frac{\partial x(t)}{\partial x(0)}= \lim_{\Delta x \to 0} \frac{x(t,x(0)+\Delta x,p(0))-x(t,x(0),p(0))}{\Delta x}.
	\label{otoc-classical}
\end{align}
To evaluate the right-hand side of this equation, we need to compute the positions of the two particles $x(t,x(0)+\Delta x,p(0))$ and $x(t,x(0),p(0))$ (see Fig.~\ref{Fig-phase}).
Generally, we cannot obtain the particle positions explicitly, and we evaluate them numerically and extrapolate the limit $\Delta x\to 0$.

\begin{figure}
	\begin{center}	
		\begin{tabular}{cc}
			\begin{minipage}{0.5\hsize}
				\begin{center}
					\includegraphics[scale=0.6]{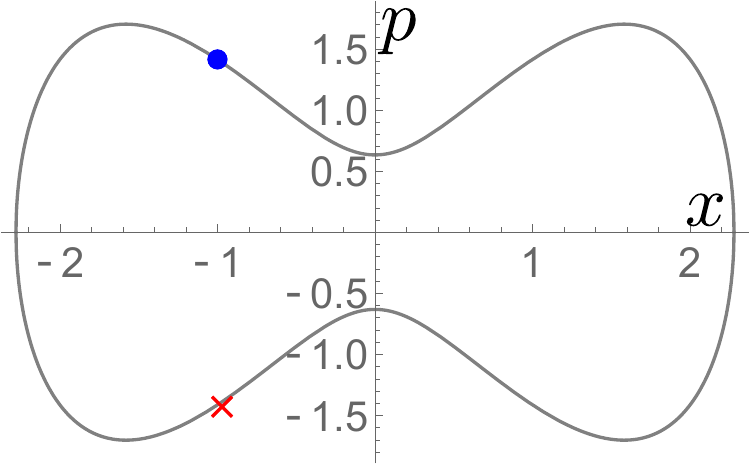}\\
				\end{center}			
			\end{minipage}
			\begin{minipage}{0.5\hsize}
				\begin{center}
					\includegraphics[scale=0.5]{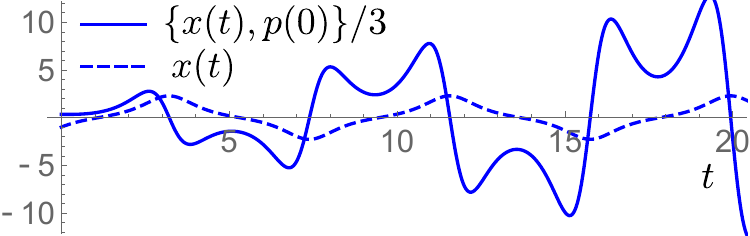}\\
					\includegraphics[scale=0.5]{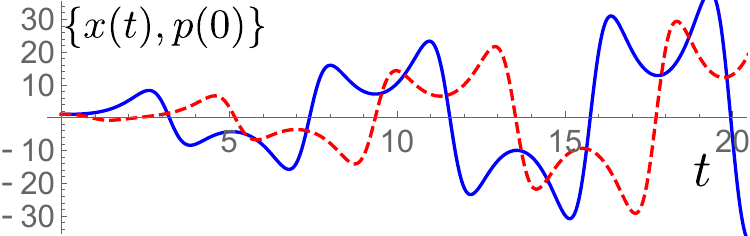}\\
				\end{center}
			\end{minipage}
		\end{tabular}
		\caption{ We show $\{x(t),p(0)\}$ for two classical particles in the potential $V(x)=-ax^2+b x^4$ with the same energy $(E=0.2)$.
			Left panel: the two particles in phase space at $t=0$. The solid line denotes the constant energy curve.
			Right panel (top): particle position $x(t)$ (blue dashed line) and $\{x(t),p(0)\}$ (blue solid line) for the blue dot in the left panel.
			Here, $x(t)$ shows a periodic motion with a period $T\simeq 8.4$.
			Correspondingly, $\{x(t),p(0)\}$ is roughly periodic but progressively increases.
			It shows the exponential development when $x(t)$ passes $0$, which is the position of the potential hill, and it takes the maximum value before the turning point ($x \simeq 2.3$) and suddenly decreases around the turning point.
			Right panel (bottom): $\{x(t),p(0)\}$ for the blue dot (the blue line) and for the red cross (the red dashed line) in the left panel.
		}
		\label{Fig-phase-2pt}
	\end{center}
\end{figure}

Let us consider the motions of the two particles in \eqref{otoc-classical} in the potential $V(x)=-ax^2+bx^4$.
Each particle periodically moves in phase space.
The period depends on the energy of the particle, and it becomes longer as $E \to E_{\rm cr}$.
(Actually, it diverges at $E=E_{\rm cr}$, and its motion is not periodic anymore.)
Thus, for example, in the case of the two particles plotted in Fig.~\ref{Fig-phase}, 
the period of the left particle is longer, and, when the left particle returns to its original position, the right particle moves slightly ahead of its original position.
Hence, $\{ x(t),p(0) \}$ is almost periodic but progressively increases for each period (see Fig.~\ref{Fig-phase-2pt}).
In addition, since the two particle motions are almost periodic, $\{ x(t),p(0) \}$ has to be zero at least twice for each period, and
the sign of $\{ x(t),p(0) \}$ changes when it crosses these zero points. 
(Otherwise, the two particles could not return to their original positions.)
In the case of the HO \eqref{OTOC-HO}, $t=\pi/2\omega $ and $3\pi/2\omega $ correspond to the zero points.

Besides, when the particles pass near the hill of the potential, they show exponential growth similar to \eqref{classical-motion}, and $\{ x(t),p(0) \}$ also develops exponentially.
On the other hand, when the particles move in the potential valleys, their motions are like oscillators and $\{ x(t),p(0) \}$  will show a cos-type behavior similar to \eqref{OTOC-HO}.
In this way, we can roughly explain the time evolution of $\{ x(t),p(0) \}$ in Fig.~\ref{Fig-phase-2pt}.

So far, we have discussed the properties of  $\{ x(t),p(0) \}$ for a single particle.
Now, we argue $\{ x(t),p(0) \}$ for the energy eigenstate by taking an average of these single particle results, 
and we explain the behaviors shown in Fig.~\ref{Fig-ene} in quantum mechanics.
We first discuss why the Lyapunov exponents are smaller than $\lambda_{\rm saddle}$.
When we take the average, the maximum of $\{ x(t),p(0) \}$ dominates.
As we can see in Fig.~\ref{Fig-phase-2pt}, the maximum appears after the exponential developments terminate, and there, the growth is much slow.
Hence, the Lyapunov exponents of the energy eigenstates are always smaller than $\lambda_{\rm saddle}$.\footnote{
	In Fig.~\ref{Fig-phase-2pt}, $\{ x(t),p(0) \}$ of each particle grow larger and larger for late times, but they are incoherent.
	This may explain the slow growth of $ -\langle [x(t),p(0)]^2 \rangle $ at late times shown in Fig.~\ref{Fig-ene}.  
	}

Next, we discuss why we do not observe clear exponential developments in $\langle [x(t),p(0)] \rangle $.
This is because $\{ x(t),p(0) \}$, for single particles, take positive and negative values, which may cancel each other out when we take the average.
This is very different from the OTOC $\langle [x(t),p(0)] \rangle $ for the wave packets, where we have seen clear exponential developments as in Fig.~\ref{Fig-wave}.
On the other hand, at the critical energy $E = E_{\rm cr}$, the particles near the saddle points dominate, and such cancellations are suppressed.
Hence, we observe the exponential growth as in Fig.~\ref{Fig-ene-cr}.

\section{Discussions}
\label{sec-discussion}

We have studied the OTOC $\langle [x(t),p(0)]^n \rangle $ in the IHO.
They have a peculiar property of not showing any quantum fluctuations independent of the quantum states.
This suggests that the OTOCs can be regarded as ideal indicators of the butterfly effect in the IHO.
Hence, we expect that the IHO may work as a starting point of perturbative calculations in unstable systems and chaos, and the Lyapunov exponents
of the systems may be extracted through the OTOCs.
(The situation may be analogous to HOs in stable systems.)
Indeed, we have seen that we can derive the Lyapunov exponents of the saddle points by preparing sufficiently localized wave packets or by employing an energy eigenstate and taking the limit $E \to E_{\rm cr}$ and $\hbar \to 0$ in one-dimensional quantum mechanics.
We have also shown that the properties of the OTOCs of the energy eigenstates can be understood through classical mechanics.
However, one-dimensional system is integrable, and it is important to apply our results to more complicated systems or genuine chaotic systems.
See Refs.~\cite{Xu:2019lhc, Bhattacharyya:2020art, Hashimoto:2020xfr, Romatschke:2020qfr, Hashimoto:2017oit, Rozenbaum:2017zfo, Rozenbaum:2019kdl, Chavez-Carlos:2018ijc, Zhuang:2019jyq, Prakash:2020fsj, Prakash:2019kip, Akutagawa:2020qbj, Bhagat:2020pcd, Kidd:2020mtu} and references therein for investigations of OTOCs in few-body quantum mechanics.

In addition, as we mentioned in the Introduction, IHOs have a wide range of applications from condensed matter systems to quantum gravity and string theories.
Thus, it is interesting to study the implications that the OTOCs do not receive any quantum corrections in these systems.

\paragraph{Acknowledgements}
The author would like to thank Koji Hashimoto and  Ryota Watanabe for valuable discussions and comments.
Part of the numerical computation in this work was carried out at the Yukawa Institute Computer Facility.
The work of T.~M. is supported in part by Grant-in-Aid for Scientific Research C (No. 20K03946) from JSPS.

 \bibliographystyle{unsrt}
 \bibliography{AHR}

\end{document}